\documentstyle[11pt,paspconf]{article}

\input epsf
\newdimen\fighsize \def\epsscale#1{\fighsize=#1\hsize} \epsscale{1}
\def\plotone#1{\epsfxsize=\fighsize\centerline{\epsfbox{#1}}}
\def\plottwo#1#2{\centerline{\epsfxsize=.5\fighsize\epsfbox{#1}\hss
                             \epsfxsize=.5\fighsize\epsfbox{#2}}}
\font\bmit=cmmib10
\def\bbeta{\hbox{\bmit\char"0C}}
\def\btheta{\hbox{\bmit\char"12}}
\def\bthim{\btheta_{\rm image}}
\def\Dl{D_{\rm l}} \def\Ds{D_{\rm s}} \def\Dls{D_{\rm ls}}
\def\zl{z_{\rm l}}

\def\<#1>{\langle\hbox{#1}\rangle}
\def\half{\hbox{$1\over2$}}

\begin{document}

\title{Cluster Reconstruction from Combined Strong and Weak Lensing}

\author{Prasenjit Saha}
\affil{Department of Physics, University of Oxford, UK}

\author{Liliya L.R. Williams}
\affil{Department of Physics and Astronomy, Univ. of Victoria, BC, Canada}

\author{Hanadi M. AbdelSalam}
\affil{Kapteyn Sterrewacht, Rijksuniversitaet Groningen, The Netherlands}

\begin{abstract}
The lensing information provided by multiple images, arclets, and
statistical distortions can all be formulated as linear constraints on
the arrival-time surface, and hence on the mass distribution.  This
reduces cluster lens reconstruction from combined strong and weak
lensing to a standard type of inversion problem.  Moreover, the mass
sheet degeneracy is broken if there are sources at different
redshifts.

This paper presents a reconstruction technique based on these ideas,
and a mass map and discussion of Cl\thinspace 1358+62.
\end{abstract}

\keywords{cosmology: gravitational lensing, galaxies: clusters:
individual: Cl1358+62}

\section{Introduction}
The different regimes of cluster lensing---the inner region where
multiple images, including giant arcs are found, further out where
highly sheared but singly imaged arclets are found, and the outer
regions showing statistical shear---have in the past been approached
with quite different modeling methods. But these apparently separate
regimes can be studied in a unified way.  The key is to express
lensing information in terms of the arrival surface.

In earlier work (see AbdelSalam et al.\ 1998) we combined the
multiple-image and arclet regimes and argued that extending to
statistical shear was a straightforward algorithmic issue.  This paper
makes that extension.

\section{The arrival-time surface}
The creature we will mostly be concerned with is the scaled
arrival-time surface $\tau(\btheta)$, and it is expressed as follows:
\begin{eqnarray}
   \label{green-eq}
   \tau(\btheta) & = & \half(\btheta-\bbeta)^2 - {(\Dls/\Ds)\over\pi}
       \int\!\ln|\btheta-\btheta'|\,\kappa(\btheta')\,d^2\btheta' \\
   \label{oper-eq}
   & = & \half(\btheta-\bbeta)^2 - 2(\Dls/\Ds)\;\nabla^{-2}\kappa(\btheta).
\end{eqnarray}
As usual, $\btheta$ and $\bbeta$ are angular locations on the image
and source planes respectively, and the $D$'s are angular diameter
distances in units of $c/H_0$.  However, $\kappa$ is the convergence
{\it for sources at infinity,} (hence the factors of $\Dls/\Ds$) to
make dealing with multiple source redshifts more convenient.  The form
of Eq.\ (\ref{green-eq}) is then familiar, while Eq.\ (\ref{oper-eq})
is just a shorthand for the same thing with $\nabla^{-2}$ denoting an
`inverse Laplacian operator' in two dimensions.

To get back to physical arrival time and surface density we use
\begin{eqnarray}
   \hbox{\rm time} & = & h^{-1}\tilde z \times
        {\rm 80\,days\,arcsec^{-2}} \times\tau(\btheta) \\
   \hbox{surface density} & = & h^{-1}\tilde z\times
     1.2 \times {10^{11}M_\odot\rm\,arcsec^{-2}} \times\kappa(\btheta)
\end{eqnarray}
where $\tilde z=(1+\zl)\Dl$ but is $\simeq \zl$.

Lensing data constrain the arrival-time surface and hence the mass
distribution $\kappa(\btheta)$ in various ways.  We can identify three
types of constraints: on the values of the arrival time surface at
particular points, on the first derivative, and on the second
derivative.

The first type is when we know the height difference between two or
more points on the arrival-time surface.  This is all-important in
time-delay quasars and $H_0$ measurements, because time delays supply
just this height-difference information.  But it is not relevant to
cluster lensing, at least not yet, though we can always hope for a
supernova in a giant arc.

The second type of constraint comes from stating Fermat's principle at
each image location, i.e., 
\begin{equation}
\nabla\tau(\bthim) = 0.
\label{grad-eq}
\end{equation}
Since image positions $\bthim$ can generally be measured very
accurately, the Eq.\ (\ref{grad-eq}) tells us that the arrival-time
surface has zero gradient at some known point.  This type of
constraint is only useful if we have multiple images; in that case we
have $2\<images>$ equation of the form (\ref{grad-eq}) but only two
unknown source coordinates to solve for, giving us some net
constraints on $\kappa(\btheta)$.  In general there are
$2(\<images>-\<sources>)$ constraints.

\begin{figure}
\epsscale{0.3}
\plotone{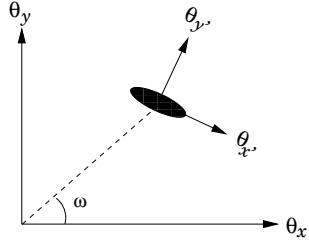}
\caption{Schematic of an arclet elongated with position angle
$\omega$, with $(\theta_{x'},\theta_{y'})$ being a coordinate system
aligned with the elongation.}
\label{fig_axes}
\end{figure}

The third type of constraint involves the curvature or second
derivative of the arrival-time surface, and may come from either
arclets or from statistical distortions.

Consider first a situation like Fig.\ \ref{fig_axes}, where an arclet
is observed elongated with position angle $\omega$.  If we are
confident that the arclet has been elongated by a factor of at least
$k$, we can consider the coordinate system $(\theta_{x'},\theta_{y'})$
aligned with the elongation and write
\begin{equation}
k    \left|{\partial^2\over\partial\theta_{x'}^2}\tau(\bthim)\right|
\leq \left|{\partial^2\over\partial\theta_{y'}^2}\tau(\bthim)\right|.
\label{axes-eq}
\end{equation}
Further, since $\omega$ is known, we can transform Eq. (\ref{axes-eq})
to the unrotated $(\theta_x,\theta_y)$ coordinates.  It is not
necessary to have highly accurate values of the elongation or its
orientation $\omega$---all we have to do is set $k$ conservatively
enough that the inequality (\ref{axes-eq}) is valid.  The arclet used
may itself be part of a multiple-image system; if so we will have to
guess the parity to remove the absolute value signs in
(\ref{axes-eq}).

The same idea can be applied to statistical distortion.  In this case
the data provide an estimate and uncertainty
\begin{equation}
{\gamma_i\over1-\kappa} = g_i\pm\Delta g_i,
\end{equation}
where $\gamma_i$ and $\kappa$ are components of the second derivative
of $\tau(\btheta)$.  We can rewrite this in various ways, such as
\begin{equation}
(1-\kappa(\bthim))g_i - \Delta g_i \leq \gamma(\bthim) \leq
(1-\kappa(\bthim))g_i + \Delta g_i.
\label{stat-eq}
\end{equation}
(We have assumed $\kappa\Delta g_i$ is negligible here.)

A key feature of the constraint equations (\ref{grad-eq}),
(\ref{axes-eq}), and (\ref{stat-eq}) is that they are all linear in
the unknowns $\bbeta$ and $\kappa(\btheta)$.  Which is to
say\footnote{With one exception: scalar magnification data are
quadratic in $\kappa$ and so (\ref{matr-eq}) does not apply. Taylor
and collaborators (see e.g., Dye \& Taylor 1998) have developed
reconstruction techniques for this case.  However, if magnification
information is present along with shear information, together they
give (at least in principle) the tensor magnification which is linear
in $\kappa$, and so (\ref{matr-eq}) again applies.}
\begin{equation}
   {\textfont1=\tenrm
    \pmatrix{Lensing \cr data} = 
    \pmatrix{A & messy & but & linear & operator\cr
             also & involving & the & same & data }
    \pmatrix{The \cr lens's \cr projected \cr mass \cr distribution \cr}
   }
\label{matr-eq}
\end{equation}
and moreover this is true for all lensing regimes: multiple-image, arclet,
and statistical shear.

\section{Mass reconstruction}
Eq.\ (\ref{matr-eq}) summarizes the advantage of casting the
observational information as constraints on the arrival-time surface.
Lens reconstruction is reduced to a linear inversion problem.  It is
(as symbolized by the $2\times5$ matrix) a highly underdetermined
problem, so in order to reconstruct the lens it is necessary to add
extra information (sometimes called a prior). Also as indicated in the
$2\times5$ matrix, the data enter into the linear operator---an
unusual complication.  However, as we discussed above, the data enter
there either as image positions which are known very accurately, or as
inequalities which can be set conservatively; so this issue does not
introduce new difficulties.

At least four different possibilities for lens reconstruction now
suggest themselves
\begin{itemize}
\item Put $\tau(\btheta)$ on a grid and regularize.  This may be the
simplest approach.
\item Use basis function expansions for $\kappa(\btheta)$ and
$\tau(\btheta)$. A Fourier-Bessel expansion
$$ \tau(\btheta) = \sum_{mn} c_{mn} J_m(k_{mn}\theta) \exp(im\phi) $$
is particularly attractive as it would trivially relate $\tau$ and
$\kappa$.
\item Pixellate $\kappa(\btheta)$ and regularize.  This is the one we
have implemented.  The mass is distributed on square tiles, each
having constant but adjustable density, and $\tau(\btheta)$ comes from
computing Eq.\ (\ref{green-eq}) exactly.  Note that although the mass
distribution is discontinuous, the arrival-time surface is smooth.
Unlike in the two previous possible approaches, pixellating the mass
makes it is easy to enforce non-negativity of the mass distribution.
\item Pixellate $\kappa(\btheta)$ and use maximum entropy. See Bridle
et al.\ (this volume) for an example of this in a somewhat different
context.
\end{itemize}

\begin{figure}
\epsscale{0.5} \plotone{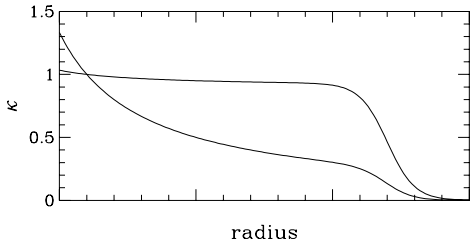}
\caption{At fixed redshift, the mass sheet degeneracy leaves
$1-\kappa$ uncertain by a multiplicative factor, while the monopole
part of $\kappa$ is undetermined outside the region of the images.  As
a result, the two mass profiles shown will produce identical
multiple-image and shear data out to the cliff.}
\label{fig_msd}
\end{figure}

The regularization we applied is to minimize
\begin{equation}
\int\left(\kappa - {\<$\kappa$>\over\<$L$>}L\right)^2 \,d^2\btheta
         + \epsilon^4 \int (\nabla^2\kappa)^2 \, d^2\btheta
\label{reg-eq}
\end{equation}
while of course enforcing all the lensing constraints. The first term
in (\ref{reg-eq}) tends to minimize mass-to-light variation, since 
$\<$\kappa$>/\<$L$>$ is a mean $M/L$.  The second term tends to
smooth, with $\epsilon$ a sort of smoothing scale.  By regularizing
with respect to different light distributions we can get an estimate
of the uncertainty.

Since the regularizing functional (\ref{reg-eq}) is quadratic in
$\kappa$, minimizing it subject to the linear lensing constraints is
readily implemented through quadratic programming algorithms.  The
disadvantage is that the storage needed is $\simeq2\<number of
pixels>^2$ limiting us to of order 5000 pixels.  The solution is to
have adaptive pixellation (much like tree codes in $N$-body
simulations), with smaller pixels for the inner parts of the cluster
and large pixels outside.


\section{The mass-sheet degeneracy}
After enthusing about how easy lens reconstruction is, it is well to
make a cautionary remark about the main source of uncertainty.  Here
again, the arrival-time surface is very useful.

\begin{figure}
\plottwo{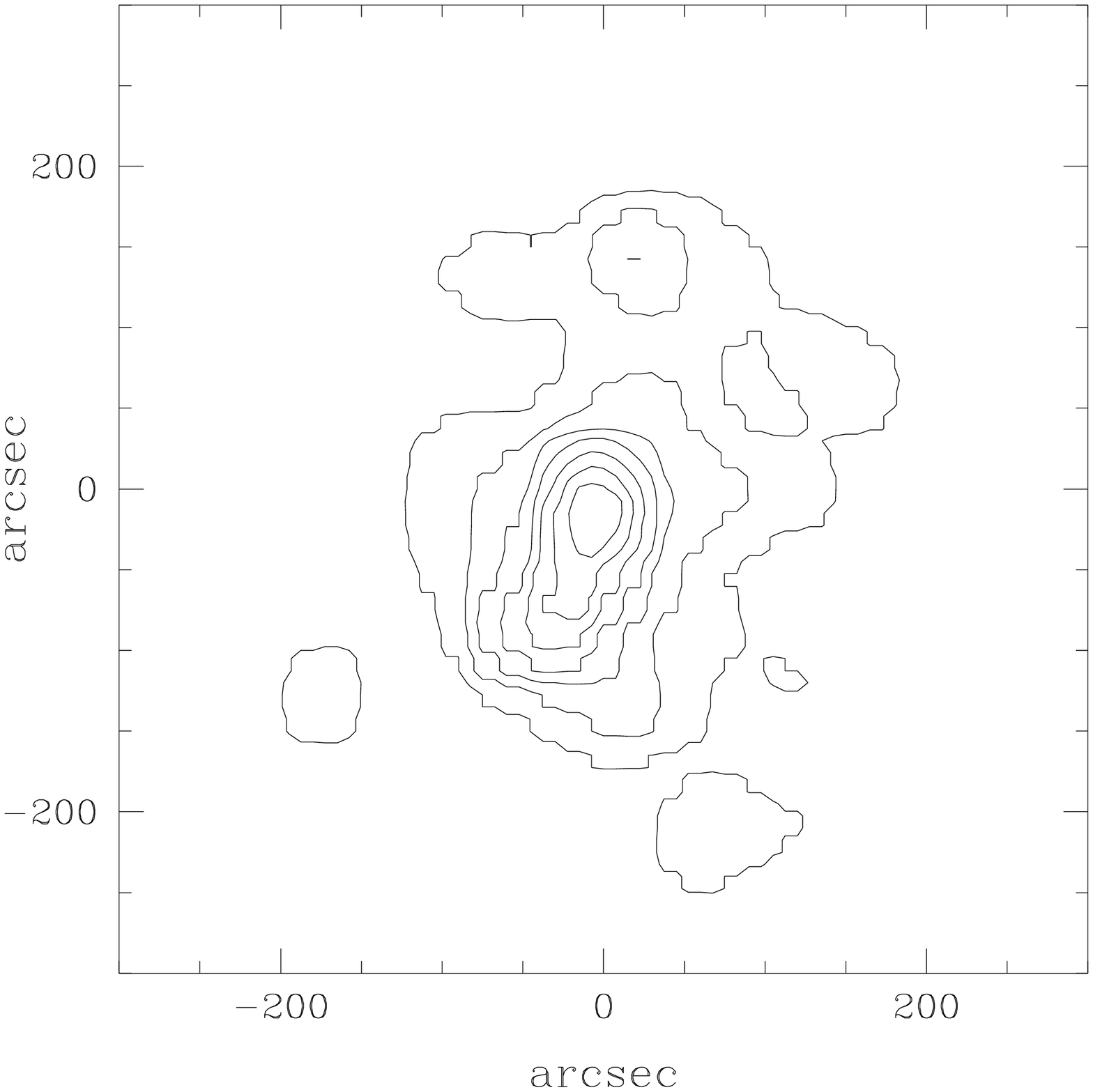}{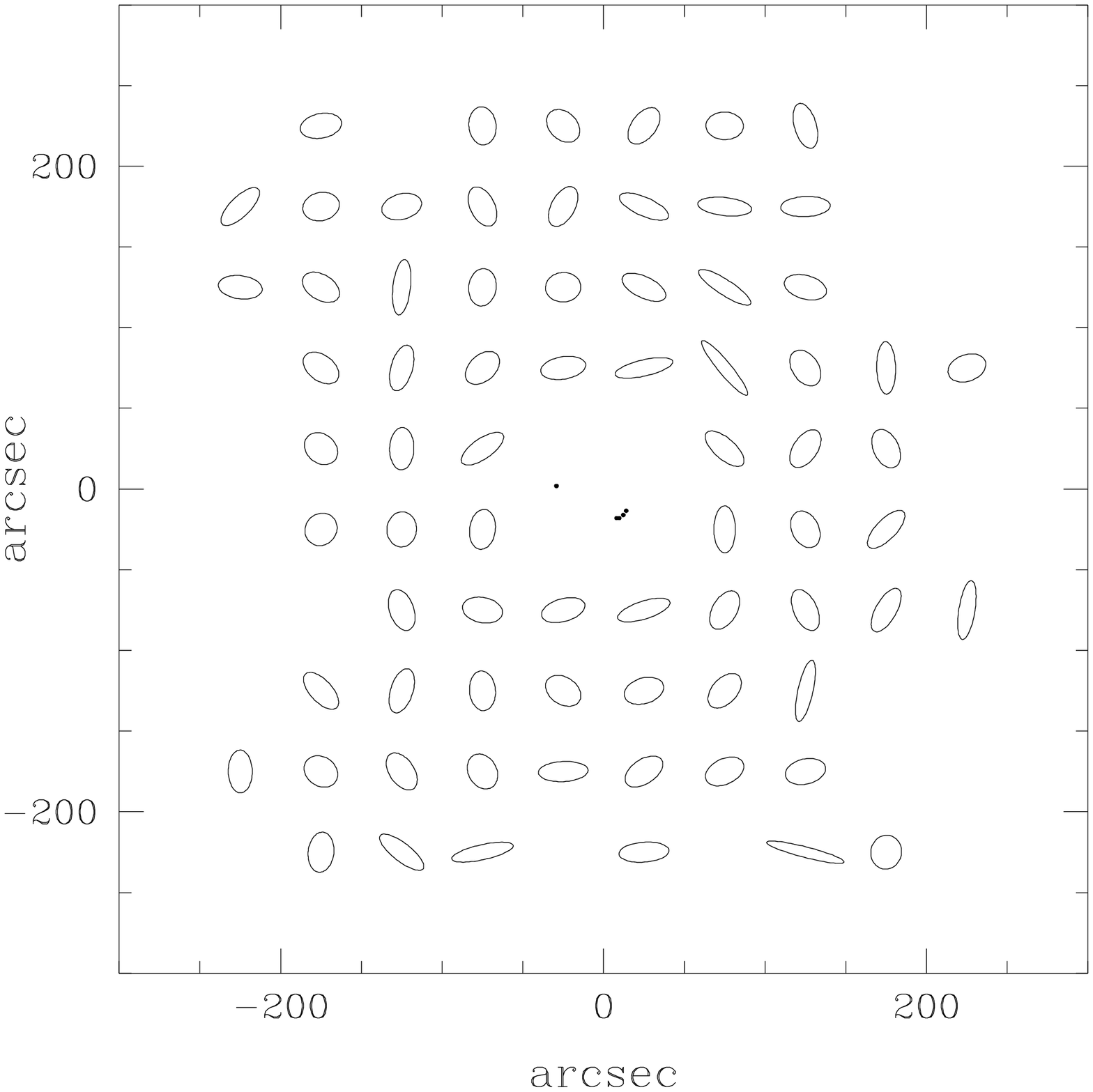}
\caption{{\bf Left:} Cluster luminosity contours in Cl\thinspace
1358+62. from the catalog of galaxy $V_z$'s in van Dokkum et al
(1998).  Likened at the conference to a left facing seahorse with a
ponytail. {\bf Right:} Shear map (exaggerated by a factor of 5) from
Hoekstra et al (1998).  We have excised the central region to avoid
averaging the shear where it is varying rapidly.  Also note that
unlike Fig.~14 in Hoekstra et al., we have binned without smoothing
here.}
\label{fig_data}
\end{figure}

If we multiply $\tau(\btheta)$ by a constant factor, we just stretch
the surface vertically, and geometrically it is clear that neither
image locations nor their relative magnifications change.  More
formally, we can rearrange Eq.\ (\ref{oper-eq}) and drop the
irrelevant term $\half\bbeta^2$ to get
\begin{equation}
\tau(\btheta) = 2\nabla^{-2}\left(1-{\Dls\over\Ds}\kappa\right)
              - \half\btheta\cdot\bbeta.
\end{equation}
If we now multiply both $\left(1-{\Dls/\Ds}\kappa\right)$ and $\bbeta$
by some constant $a$, the image structure will be unchanged; only time
delays and total magnification will get multiplied by $a$ and $a^{-1}$
respectively (the latter because rescaling $\bbeta$ rescales all the
sources).  Therefore lens reconstruction from image structure (without
absolute magnifications) leaves $\left(1-{\Dls/\Ds}\kappa\right)$
uncertain by a constant factor. This is the mass-sheet degeneracy.

It may appear at first that the mass-sheet degeneracy can be
eliminated by requiring $\kappa$ to vanish at large distances.  But
outside the observed region, the monopole part of $\kappa$ is
completely unconstrained.  So the mass-sheet degeneracy is equally
effective with a mass disk larger than the observed region.  Figure
\ref{fig_msd} illustrates.

\section{Reconstruction of Cl\thinspace 1358+62 (The Seahorse Cluster)}
This cluster seemed the obvious test case for combined strong and weak
lensing reconstruction in a relatively large field.  The cluster
itself is at $z=0.33$ and its inner part lenses a $z=4.92$ galaxy into
a red arc; this was identified by Franx et al.\ (1997) who also
presented a strong lens model.  Hoekstra et al.\ (1998) measured the
shear field in an HST WFPC2 mosaic and reconstructed a mass map from
weak lensing.  The present work combines both regimes.

\begin{figure}
\plottwo{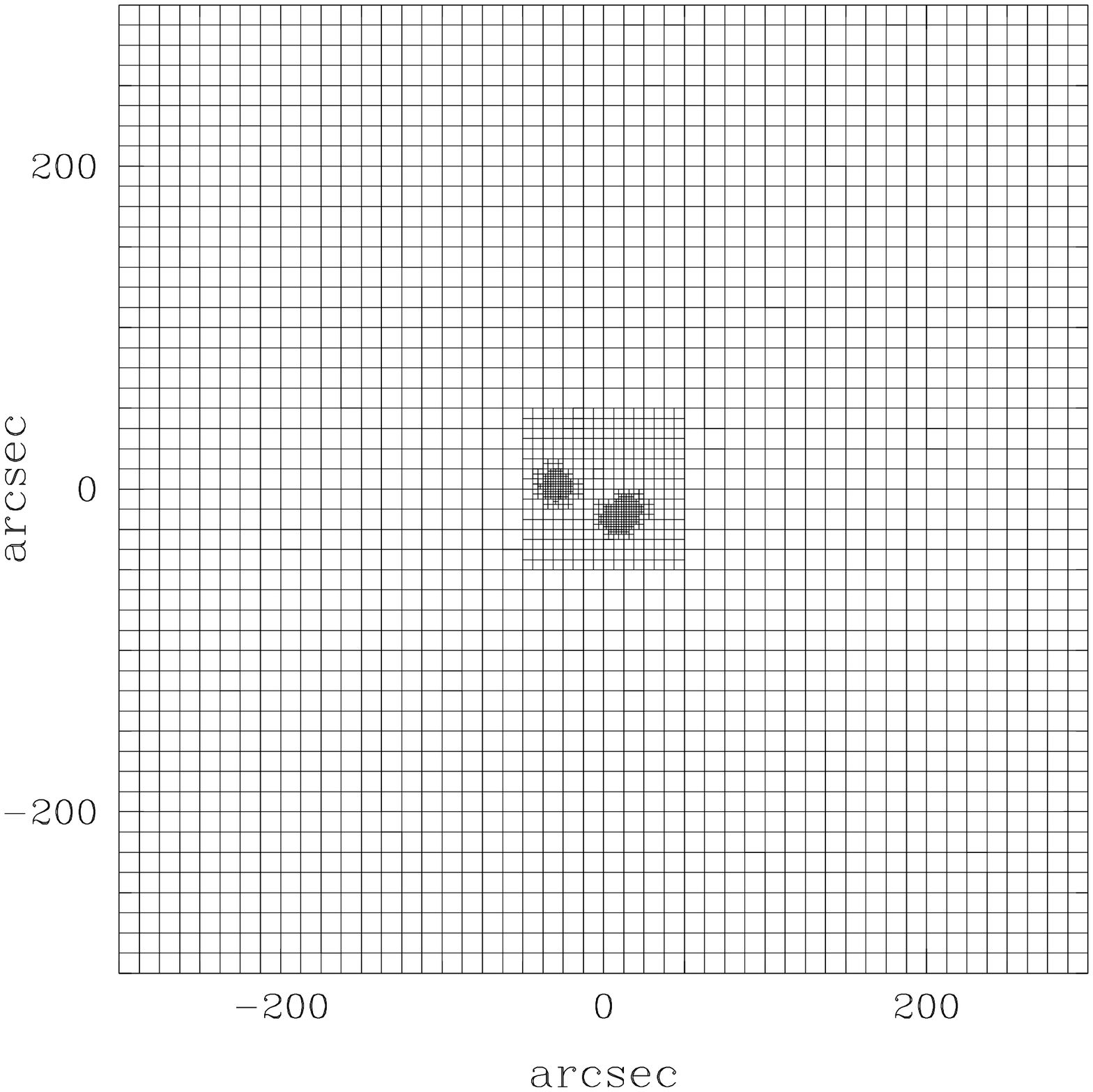}{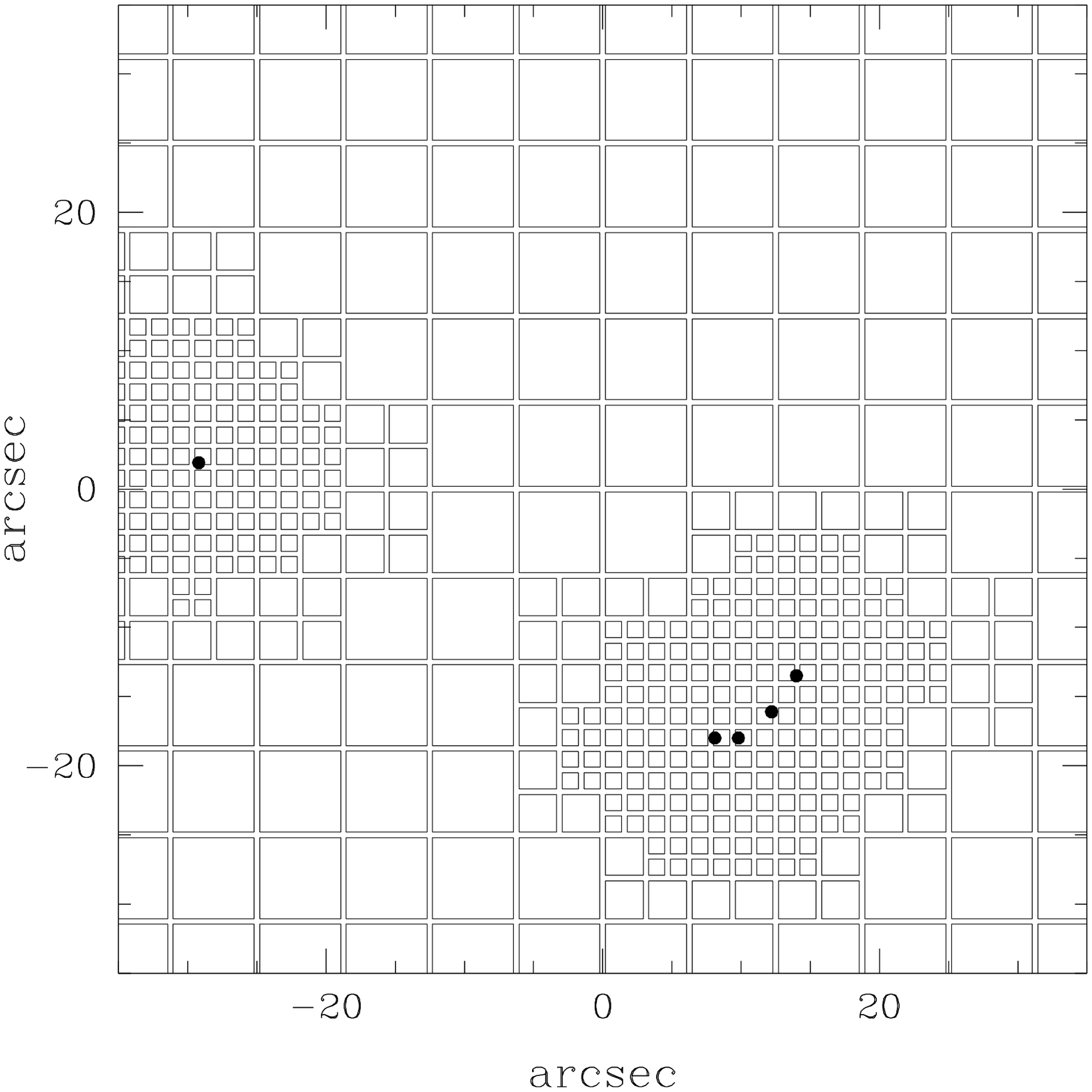}
\caption{Pixellation used for the mass reconstruction of the Seahorse
cluster. Note that is completely adaptive, and allows us to increase
the resolution at any desired place---in this case the region around
the red arc.  {\bf Left:} The full field. {\bf Right:} The inner
region, with dots marking features in the red arc.}
\label{fig_pix}
\end{figure}

Figure \ref{fig_data} illustrates the data on this cluster that we
have used: a shear field derived by binning the individual
polarization measurements kindly provided by Hoekstra and
collaborators and assuming a constant $z=1$ for the background
galaxies, and a smoothed $V_z$ light distribution.  Figure
\ref{fig_pix} shows the pixellation we used.

\begin{figure}
\plottwo{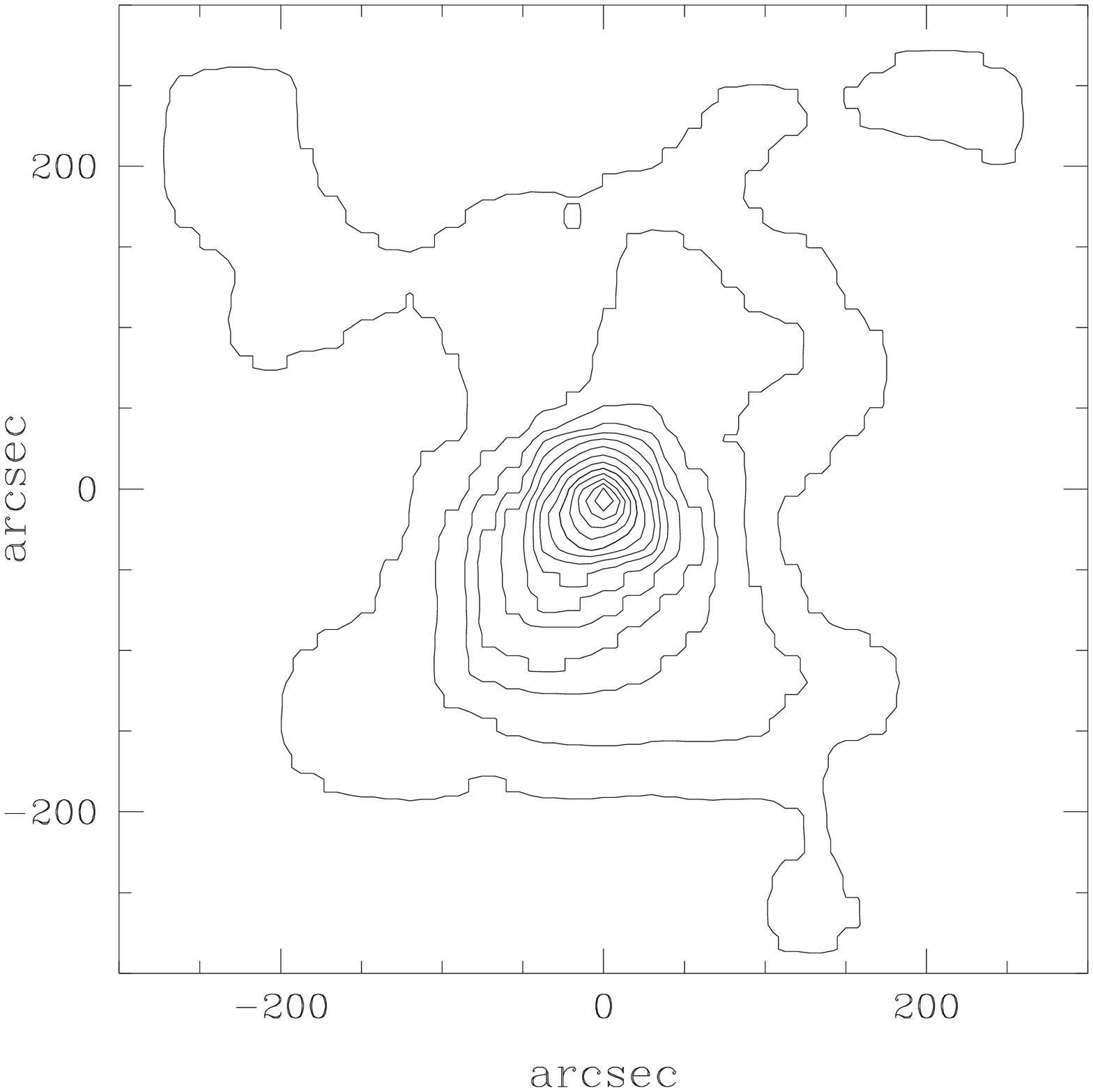}{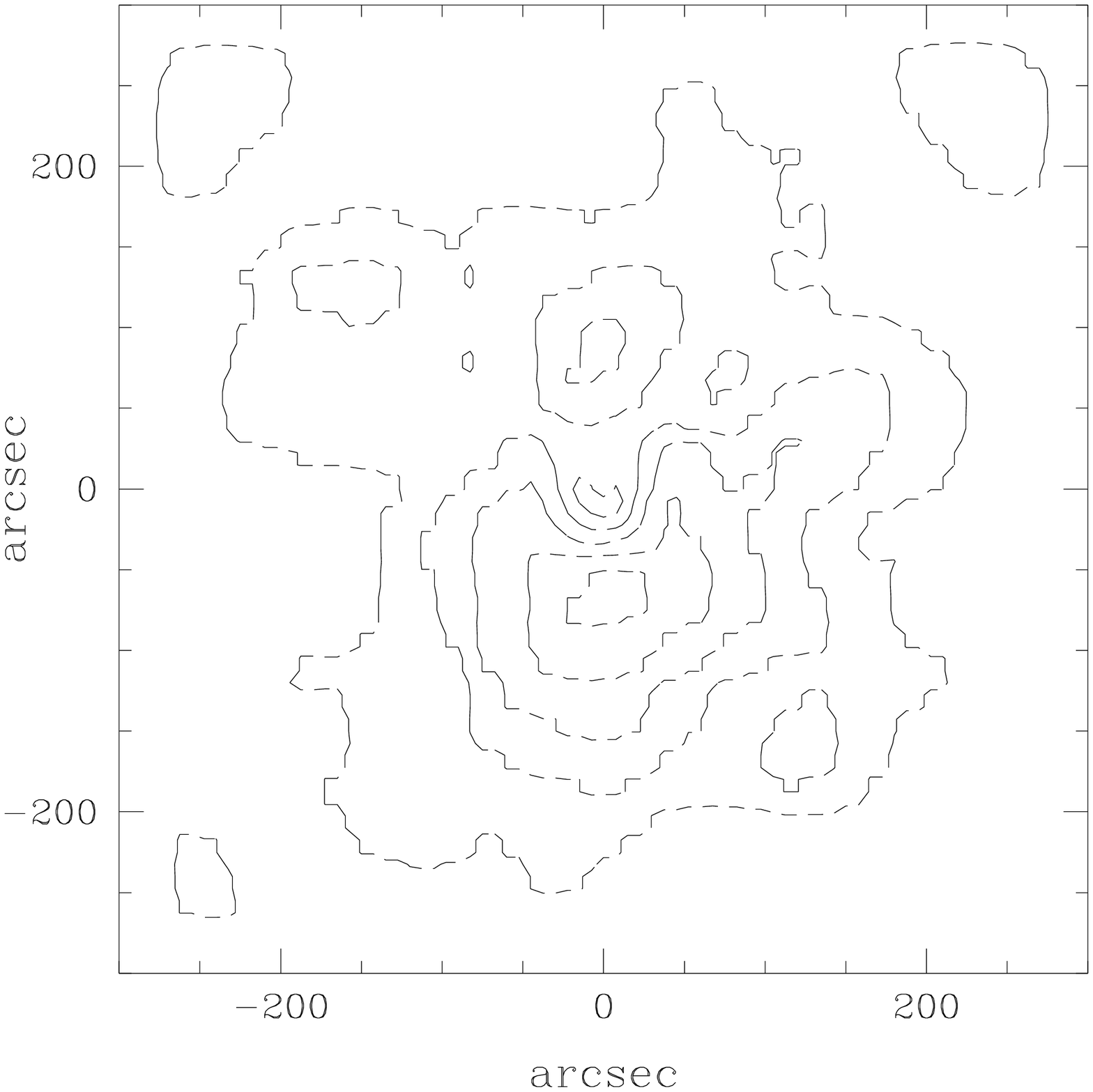}
\caption{Mass map of the Seahorse cluster. {\bf Left:} Contours of
$\kappa$ in steps of 0.1. {\bf Right:} Contours of $\Delta\kappa$ in
steps of 0.025.}
\label{fig_mass}
\end{figure}

Figure \ref{fig_mass} is our mass map, computed by regularizing with
respect to the light and a smoothing scale $\epsilon$ changing from
$\simeq5''$ in the center to $\simeq1'$ at the edge.  Our estimated
uncertainty is derived from an ensemble of reconstructions where we
rotated the light map by arbitrary angles and shifted it randomly by
up to $100''$ and regularized with respect to these altered light
maps.  The mass map resembles Fig.\ 15 of Hoekstra et al.\ (1998) but
tends to be smoother.  Also the overall normalization is somewhat
higher; this is probably due to different treatments of the boundary,
though with the mass-sheet degeneracy.  (The red arc at $z=4.92$
compared to $z=1$ assumed for the weakly lensed galaxies, reduces the
effect of the mass-sheet degeneracy, but does not eliminate it.)  Our
central density is higher, but this is as expected since inclusion of
a multiply-imaged system immediately forces $\kappa>1$.  There is some
indication that the mass peak is offset (by some 10s of kpc) to the
south of the light peak, a natural thing to expect if the cluster is
asymmetric and galaxy formation is biased.  But this offset is
tentative; in Abell~2218, where the inner region is much richer in
lensing and better constrained, the evidence for an offset is more
compelling (AbdelSalam et al.\ 1998).

\begin{figure}
\plottwo{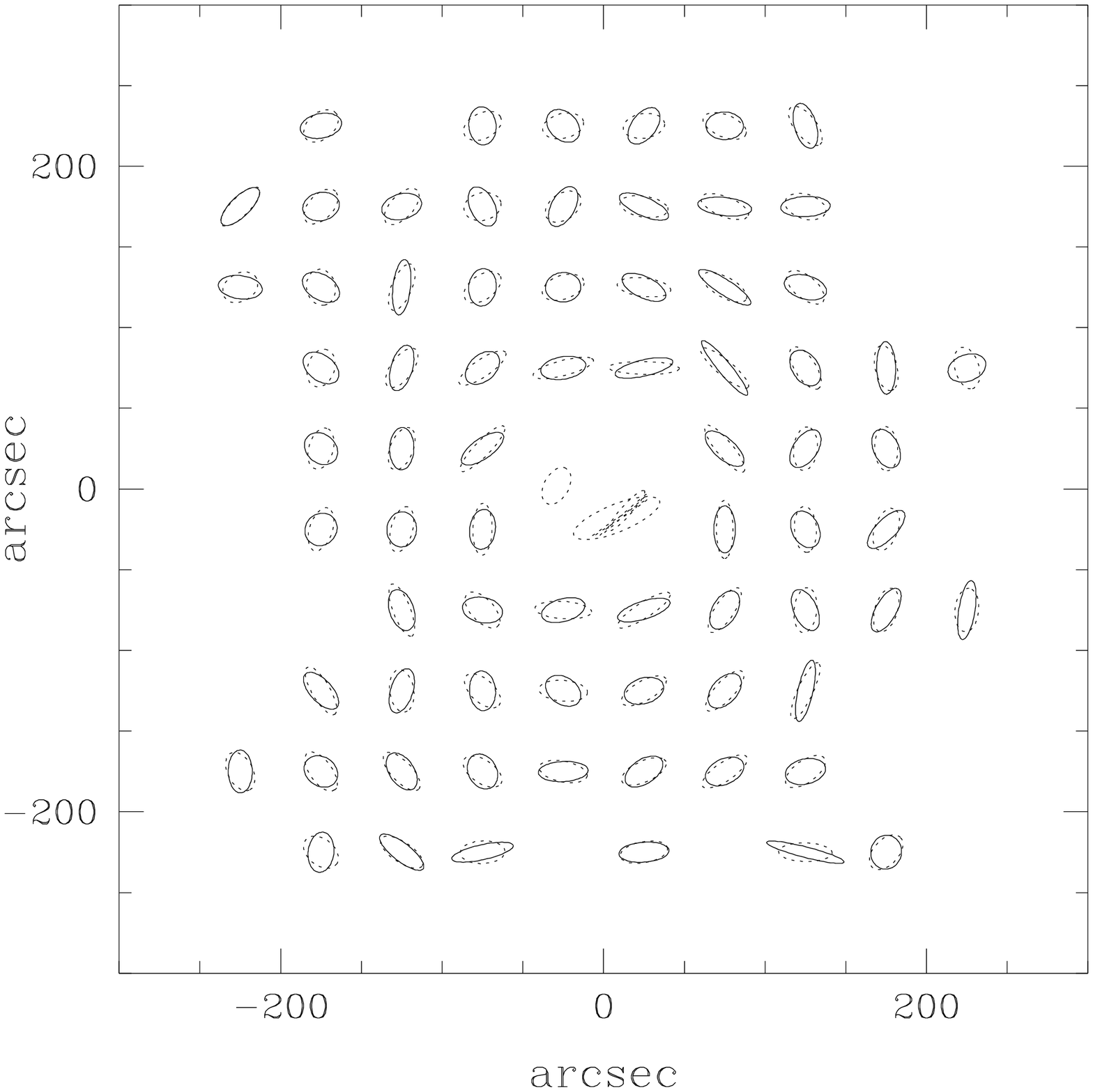}{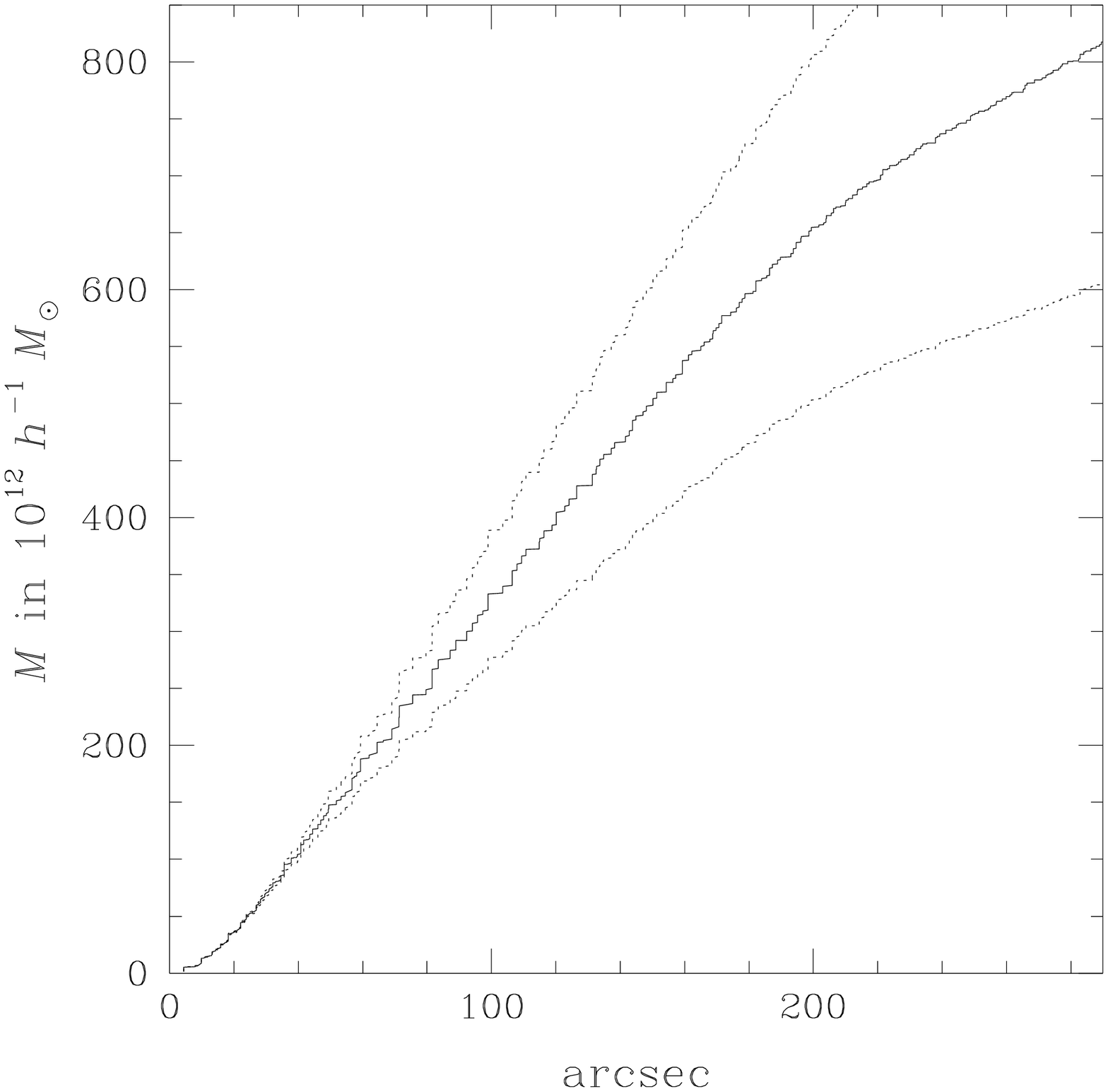}
\caption{{\bf Left:} Shear map. The superimposed ellipses show the
observed and reconstructed (dotted) shear, both exaggerated by a
factor of five. As we might expect, the reconstructed shear appears
more correlated than the observed. The ellipses in the inner region
show the reconstructed shear (not exaggerated) at the locations of the
red arc.  {\bf Right:} Enclosed mass and estimated uncertainty.}
\label{fig_rec}
\end{figure}

Figure \ref{fig_rec} shows the enclosed mass out to different radii
and the reconstructed shear.  It is surprising that the enclosed mass
after angular averaging, looks so `isothermal' even though the mass
map is very asymmetric.  We obtain $M=(10\pm1)\times
10^{14}M_\odot\rm\,Mpc^{-1}$, corresponding to a formal Einstein
radius of $43''$ and a formal isothermal los dispersion of
$(990\pm70)\,\rm km\,sec^{-1}$.  The estimated mass-to-light is
$(380\pm60) h\,M_\odot L_{\odot V}^{-1}$.  The CNOC survey (Carlberg
et al.\ 1997) measured $910\,\rm km\,sec^{-1}$ for the los velocity
dispersion and estimated $M/L=229h^{-1}M_\odot L_{\odot V}^{-1}$.

\section{Discussion}
We have developed a mass reconstruction technique that deals with all
the cluster lensing regimes simultaneously, from multiple images in
the central regions to weak shear in the outer regions, and moreover
with adaptive resolution.  Code implementing this is available from
the authors.  Variants of our technique (suggested in Section 3) may
also be of interest in future work.

At this stage it appears that cluster mass reconstruction is very good
at recovering {\it features\/}---a good example are offsets between
mass and light peaks in clusters, which may be indirect evidence for
biased galaxy formation.  But the {\it calibration\/} of the mass
(even with perfect shear data) is still problematic---the mass sheet
degeneracy is a vicious effect.  We suggest that recovering the
redshift distribution and the absolute magnification of the background
galaxies (even with large uncertainties) are the best hope of breaking
this degeneracy, and is likely to be very rewarding in future work.

\acknowledgments
We are grateful to Henk Hoekstra and collaborators for supplying us
with their shear data on Cl\thinspace 1358+62.

\end{document}